\documentclass[iop]{emulateapj}
\usepackage{natbib}
\usepackage{graphicx}
\usepackage{float}
\usepackage{epsfig}
\usepackage{booktabs}
\usepackage{CJK}

\shorttitle{Planet-disk interaction at the dead zone inner boundary}
\begin{document}
\begin{CJK*}{UTF8}{gkai}

\title{Inside-Out Planet Formation. III.\\
Planet-disk interaction at the dead zone inner boundary}
\author{Xiao Hu (胡晓)\altaffilmark{1}, Zhaohuan Zhu (朱照寰)\altaffilmark{2}, Jonathan C. Tan\altaffilmark{1,3}, Sourav Chatterjee\altaffilmark{1,4}}
\altaffiltext{1}{Department of Astronomy, University of Florida, Gainesville, FL 32611}
\altaffiltext{2}{Department of Astrophysical Sciences, Princeton University, Princeton, NJ 08544}
\altaffiltext{3}{Department of Physics, University of Florida, Gainesville, FL 32611}
\altaffiltext{4}{Center for Interdisciplinary Exploration and Research in Astrophysics (CIERA), Physics and Astronomy, Northwestern University, Evanston, IL 60208}

\begin{abstract}
The {\it Kepler} mission has discovered more than 4000 exoplanet
candidates. Many of them are in systems with tightly packed inner
planets (STIPs). Inside-Out Planet Formation (IOPF) \citep{CT14} has
been proposed as a scenario to explain these systems. It involves
sequential {\it in situ} planet formation at the local pressure
maximum of a retreating dead zone inner boundary (DZIB). Pebbles
accumulate at this pressure trap, which builds up a pebble ring, and
then a planet. The planet is expected to grow in mass until it opens a
gap, which helps to both truncate pebble accretion and also induce
DZIB retreat that sets the location of formation of the next
planet. This simple scenario may be modified if the planet undergoes
significant migration from its formation location. Thus planet-disk
interactions play a crucial role in the IOPF scenario. Here we present
numerical simulations that first assess the degree of migration for
planets of various masses that are forming at the DZIB of an active
accretion disk, where the effective viscosity is undergoing a rapid
increase in the radially inward direction. We find that torques
exerted on the planet by the disk tend to trap the planet at a
location very close to the initial pressure maximum where it
formed. We then study gap opening by these planets to assess at what
mass a significant gap is created. Finally we present a simple model
for DZIB retreat due to penetration of X-rays from the star to the
disk midplane. Overall, these simulations help to quantify both the
mass scale of first, ``Vulcan,'' planet formation and the orbital
separation to the location of second planet formation.
\end{abstract}

\section{Introduction}

Since launch in 2009, {\it Kepler} has revealed more than 4000
exoplanet candidates \citep[e.g.,][]{2015ApJS..217...31M}. A large
percentage ($\gtrsim 30\%$) of these candidates are in systems with
tightly-packed inner planets (STIPs). These are systems that usually
have 3 or more detected planets of radii $\sim1-10\:R_\oplus$ and with
periods less than 100 days \citep{2012ApJ...761...92F}. There are two
main scenarios that can produce such close-in planets: (1) formation
in the outer disk followed by inward migration
\citep[e.g.,][]{2012ARA&A..50..211K,2013A&A...553L...2C,2014IAUS..299..360C};
(2) formation {\it in situ}
\citep[][hereafter CT14]{2012ApJ.751.158, 2013ApJ.775.53, 2013MNRAS.431.3444C, CT14}. 

The inward migration scenario tends to produce planets that are
trapped in orbits of low order mean motion resonances, which is not a
particular feature of STIPs
\citep{2014prpl.conf..667B,2014ApJ.790.146}. Thus it has been proposed
that the lack of resonant pile-ups might be explained by a lower
efficiency of resonance trapping or breaking of resonance by later
dynamical processes for the typically low-mass {\it Kepler}-detected
planets \citep{2014AJ....147...32G,2015ApJ...803...33C}. 

The {\it in situ} formation scenario faces the challenge of
concentrating enough solids in the inner region
\citep{2014MNRAS.440L..11R,2014ApJ...795L..15S}. For example, supply
of pebbles by radial drift may be truncated if planet formation,
perhaps initiated by the pebble streaming instability \citep{2005ApJ...620..459Y}
, occurs in the outer disk \citep{2014A&A...572A.107L, 2015Natur.524..322L,
2015A&A...582A.112B}.
{\it In situ} formation models also face the challenge of reproducing
the observed mass versus orbital radius distributions once effects of
gas on protoplanet migration during the oligarchic growth phase are
accounted for \citep{2015A&A...578A..36O}.

The Inside-Out Planet Formation (IOPF) scenario proposed by CT14 is a
new type of {\it in situ} formation model. It starts with pebble
delivery to the midplane transition region between the innermost
MRI-active zone and a nonactive ``dead zone,'' where there is a local
pressure maximum. The pebbles trapped by the pressure maximum will
build up in a ring, which then forms a protoplanet, perhaps involving
a variety of processes including streaming
\citep{2005ApJ...620..459Y}, gravitational \citep{1964ApJ...139.1217T}
and/or Rossby wave \citep{2006A&A...446L..13V} instabilities.  The
protoplanet is expected to continue its growth, especially by pebble
accretion, until it becomes massive enough to open a gap in the
disk. This gap pushes the pressure maximum outwards by a few Hill
radii thus creating a new pebble trap that is displaced from the
planet's orbit \citep[][]{2014A&A...572A..35L}, but may also allow
MRI-activation in the region beyond the planet, which could induce
further outward retreat of the DZIB.

As discussed by CT14, planetary migration, either before gap opening
(Type I) or after gap opening (Type II), may in principle alter this
scenario. However, the expectation is that the pressure maximum
associated with this inner disk transition region will also act as a
planet trap
\citep{2006ApJ...642..478M,2009ApJ...691.1764M,2012ApJ.755.74K,2014A&A...570A..75B},
so the planet will stay at its initial formation location and keep
accreting more materials, at least until the point of gap
opening. However, this needs to be confirmed for the particular disk
structure involved in IOPF and this is one of the goals of this paper.

The disk studied by \citet{2006ApJ...642..478M} was a purely passive
disk, i.e, the temperature was set only by irradiation from the
central star.  \citet{2014ApJ...797...20Z} studied planetary migration
in a disk with an inner region heated by viscous dissipation and an
outer region heated by stellar luminosity.  This structure also
contained a local pressure maximum that acted as a trap for Type I
planetary migration. In the model for IOPF, the first planet is formed
at the outer edge of the MRI active zone around the protostar (where
$T\sim 1200$~K, leading to thermal ionization of alkali metals) and
migration needs to be studied in the context of an active disk, i.e.,
heated by viscous accretion, with accretion rates $\sim
10^{-9}\:M_\odot\:{\rm yr}^{-1}$.

The magnitude of the planet mass that leads to gap opening and
potential DZIB retreat is also of crucial importance for the IOPF
theory. Gap opening has been studied by
\citet{1986ApJ...307..395L,1993prpl.conf..749L}, who proposed a
``viscous-thermal'' criterion that sets a gap opening mass.  This mass
is sensitive to the local disk viscosity. Magnetohydrodynamic
simulations with realistic turbulence have also been carried out to
study gap opening in turbulent disks
\citep{2003MNRAS.339..993N,2013ApJ...768..143Z}.

The second goal of this paper is to study gap opening and quantify the
gap opening mass in the context of the IOPF disk model, i.e., an inner
region of an active disk where the effective viscosity is undergoing a
sharp increase in the radially inward direction. These gap opening
masses can then be compared to masses of the innermost, so-called
``Vulcan'' planets. An initial comparison of the simple, analytic gap
opening mass prediction with the observed Vulcans has been carried out
by \citet[hereafter CT15, Paper II]{CT15}, who found a predicted
scaling and normalization of planet mass versus orbital radius, $r$,
of $M_p=M_G\simeq 5.0 \phi_{G,0.3} (r/0.1\:{\rm AU}) \alpha_{-3}\:
M_\Earth$, where $\phi_{G,0.3}$ is a dimensionless parameter that
indicates the fraction of the Lin-Papaloizou mass scale that is
assumed to lead to a deep enough gap to truncate planetary accretion
of pebbles and initiate DZIB retreat. The normalization of this gap
opening planet mass at given radius also depends on the
Shakura-Sunyaev viscosity parameter, $\alpha$, in the dead zone inner
boundary region, with the fiducial value of $10^{-3}$ being adopted in
the above formula. However, this value is quite uncertain and the
observed planet masses may require a somewhat lower value of $2\times
10^{-4}$ (Paper II), or a lower value of $\phi_G$.

Finally, our third goal is investigate how gap opening may induce DZIB
retreat, which then sets the location of second planet formation. Here
we will present a simple, heuristic first exploration of this process
by modeling the location of the DZIB as set by penetration of X-rays
emitted from the protostellar corona to the disk midplane beyond the
planet-induced gap. This location then sets the radius for an imposed
radial profile of viscosity that simulates the transition region from
an MRI-active inner region to an outer dead zone. Increased viscosity
leads to reduced densities that make it easier for X-rays to penetrate
further. We present example toy models of this process in which the
location of the DZIB ends up stabilizing several initial gap widths
away from the planet. Such separations can be compared to the observed
separation of orbits of innermost STIPs planets.

In \S\ref{S:analytic}, we describe an analytic model to calculate our
disk parameters. In \S\ref{S:numerical}, we describe our numerical
set-up and test the numerical simulations against the analytic model.
In \S\ref{S:migration}, we study migration of planets of various
masses located in the DZIB region. In \S\ref{S:gapopening}, we study
gap opening and DZIB retreat. In \S\ref{S:obs}, we discuss the
implications of our results for observed planets. We conclude in
\S\ref{S:conclusions}.

\section{Analytic Estimates of Disk Structure and Gap Opening Mass}
\label{S:analytic}

As in Paper I, we follow \cite{2002apa..book.....F}'s derivation of
the structure of a viscously heated accretion disk. These results will
be compared to the numerical simulations with FARGO, described
 below in \S\ref{S:numerical}.
To achieve consistency with these simulations we find we need to make
a small correction in the choice of the vertical optical depth
equation and now adopt:
\begin{eqnarray}
\tau=0.5\Sigma\kappa,
\end{eqnarray}
with the factor of 0.5 being introduced since the disk has two faces
(or equivalently a change in the definition of ``midplane''
conditions). Thus when calculating the balance between energy
dissipation and radiative cooling, the optical depth is integrated
from the midplane to each surface of the disk. This change then leads
to modest ($\lesssim40\%$) changes in the normalization of the disk
structure equations compared to CT14. For example, for disk
mass surface density we now derive:
\begin{eqnarray}
\Sigma_g&=&\frac{2^{7/5}}{3^{6/5}\pi^{3/5}}\left(\frac{\mu}{\gamma k_B}\right)^{4/5}\left(\frac{\kappa}{\sigma_{\rm SB}}\right)^{-{1}/{5}} \alpha^{-4/5} \nonumber \\
&\times&\left(Gm_*\right)^{1/5}\left(f_r\dot{m}\right)^{3/5}r^{-{3}/{5}}\\
&\rightarrow&142\gamma_{1.4}^{-4/5}\kappa_{10}^{-1/5}\alpha_{-3}^{-4/5}m_{*,1}^{1/5}(f_r\dot{m}_{-9})^{3/5}{r_{\rm AU}}^{-3/5}\: \rm g\:cm^{-2}\nonumber
\end{eqnarray}
(CT14 derived a normalization of $106\:{\rm g\:cm^{-2}}$), where
$\mu=2.33m_{\rm{H}}=3.90\times10^{-24}\:\rm{g}$ is the mean particle
mass (assuming $n_{\rm He}=0.1n_{\rm H}$), $k_B$ is Boltzmann's
constant, $\gamma \equiv 1.4\gamma_{1.4}$ is the power law exponent of
the barotropic equation of state $P=K\rho^\gamma$ where we have
normalized for $\rm H_2$ with rotational modes excited, $\sigma_{\rm
  SB}$ is Stefan-Boltzmann's constant, $m_* \equiv m_{*,1} M_\odot$ is
the stellar mass, $\kappa \equiv \kappa_{10} 10\:{\rm cm}^2\:{\rm
  g}^{-1}$ is disk opacity \citep[normalized to expected
  protoplanetary disk values, e.g.,][]{2002ApJ...564..887W},
$f_r\equiv1-\sqrt{r_*/r}$, (where $r_*$ is stellar radius), and
$\dot{m}\equiv \dot{m}_{-9} 10^{-9}\:M_\odot \: {\rm yr}^{-1}$ is the
accretion rate.

For the disk aspect ratio we find
\begin{eqnarray}
\frac{h}{r}&=&\left(\frac{3}{128}\right)^{1/10}\pi^{-{1/5}}\left(\frac{\mu}{k}\right)^{-2/5}\gamma^{-{1}/{10}}\sigma_{\rm SB}^{-{1}/{10}}
\nonumber \\
&\times&\alpha^{-{1}/{10}}\left(Gm_*\right)^{-{7}/{20}}\kappa^{1/10}\left(f_r \dot{m}\right)^{1/5}r^{{1}/{20}}\\
&\rightarrow&0.027\gamma_{1.4}^{-1/10}\kappa_{10}^{1/10}\alpha_{-3}^{-1/10}m_{*,1}^{-7/20}(f_r\dot{m}_{-9})^{1/5}{r_{\rm AU}^{1/20}},\nonumber
\end{eqnarray}
The above equations will be used to set the initial conditions of the
FARGO simulations described below.

Given this disk structure, the viscous
criterion for the gap
opening planet mass is :
\begin{equation}
M_{G}=\phi_G\frac{40\nu m_*}{r^2 \Omega_K}.
\end{equation}
Implementing our disk model, we obtain:
\begin{eqnarray}
M_{G}&=&40\phi_G \left(\frac{3}{128}\right)^{1/5}\pi^{-{2}/{5}}\left(\frac{\mu}{k}\right)^{-{4}/{5}}\gamma^{{4}/{5}}\sigma_{\rm SB}^{-{1}/{5}}
\nonumber\\
&\times&\alpha^{{4/5}}G^{-{7/10}}m_*^{{3}/{10}}\kappa^{1/5}\left(f_r \dot{m}\right)^{2/5}r^{{1}/{10}}\\
&\rightarrow&14.08\phi_{G}\gamma_{1.4}^{4/5}\kappa_{10}^{1/5}\alpha_{-3}^{4/5}m_{*,1}^{3/10}(f_r\dot{m}_{-9})^{2/5} r_{\rm AU}^{1/10}\:M_\oplus.\nonumber
\end{eqnarray}
This is a factor $0.745$ smaller than the CT14 result. 

The full set of disk structure equations with our revised
normalizations are presented in Appendix~\ref{app:equations}.

\section{Numerical Simulation Methodology}
\label{S:numerical}

We used the 2D code FARGO-ADSG \citep[]{2008ApJ...672.1054B, 2008ApJ...678..483B}, which is
built on FARGO \citep{2000A&AS..141..165M}, but with the energy
equation and disk self-gravity implemented. Self-gravity has not been
turned on in our calculation since the disk is far from gravitational
instability in our problem, which focusses on inner regions near the
star. We also implemented a simple method of radiative cooling that
uses the disk surface density to calculate the optical depth and
cooling rate as described in \citet{2012ApJ...746..110Z}.  We set up a
disk with the same mass surface density profile, flaring index,
opacity, adiabatic index, mean molecular mass and central stellar mass
as the fiducial analytic model in \S\ref{S:analytic}.

The energy equation implemented in FARGO-ADSG is
\citep{2008ApJ...672.1054B}:
\begin{equation}
\frac{\partial e}{\partial r} + \vec{\nabla}\cdot(e\vec{v})=-\left(\gamma-1\right)e\vec{\nabla}\cdot\vec{v}+Q_{+}- Q_-,
\end{equation}
where $e$ is the thermal energy per unit area, 
$\vec{v}$ is the flow velocity
and $Q_+$($Q_-$) denote heating (cooling) source terms, assumed to be
positive quantities. In a steady accretion disk, these terms can be
written as
\begin{eqnarray}
e &= &\frac{\Sigma_g T }{\gamma -1}\frac{k_B}{\mu}=Gm_*\frac{1}{\gamma-1}\Sigma_g\left(\frac{h}{r}\right)^2 r^{-1}\\
Q_+ & = & \sum\limits_{i,j}{S_{i,j}}\frac{\partial v_i}{\partial x_j}= \nu\Sigma_g \left(\frac{dv_\phi}{dr}\right)^2\\
Q_- & = & \frac{16\sigma_{\rm SB}}{3\tau}T^4
\end{eqnarray}
where $S_{i,j}$ are components of the viscous stress tensor, 

$T$ is the midplane temperature. Note
$\tau=0.5\Sigma_g\kappa$ is the optical depth from disk midplane
to the surface\footnote{During numerical testing we found an error
  in the public version of the code: it only calculates half the value
  of $\nu\Sigma_g (dv_\phi / dr)^2$.}.

First we set up an accretion disk with a constant value of $\alpha$,
i.e., without a transition zone, and a steady accretion rate of
$10^{-9}\:M_\odot\:{\rm yr}^{-1}$, simulating a range of radii from
0.02 to 0.3~AU. In all simulations we use the EVANESCENT boundary
condition. This damps the disk values (surface density, velocities and
energy density) to the initial axisymmetric disk conditions. The
damping regions are rings within the radial ranges [$r_{\rm min},1.25
  r_{\rm min}$] and [$0.84 r_{\rm max},r_{\rm max}$], where $r_{\rm
  min}(r_{\rm max})$ is the inner (outer) radius of the disk. The
damping parameter is increased from zero to a maximum from the edge of
damping zone to the edge of the disk.

Within each orbit, the damping amplitude is not a constant: the actual
damping is the calculated damping function times a coefficient, which
gradually grows from zero to one, linearly with time.

For our standard ``medium resolution (MR),'' the disk is evenly
divided by 300 sectors azimuthally and logarithmically divided by 128
sectors radially. This gives a typical grid size of
$0.002\times0.002$~AU, which is about the same as the Hill radius of a
1-$M_G$-mass planet located at 0.1~AU. We also carry out ``high
resolution (HR)'' simulations that have twice the MR resolution, and
a few test runs (of more limited time duration) at ``super-high
resolution'' (SHR) that have four times the MR resolution.

\begin{figure}[H] 
\centering
\plotone{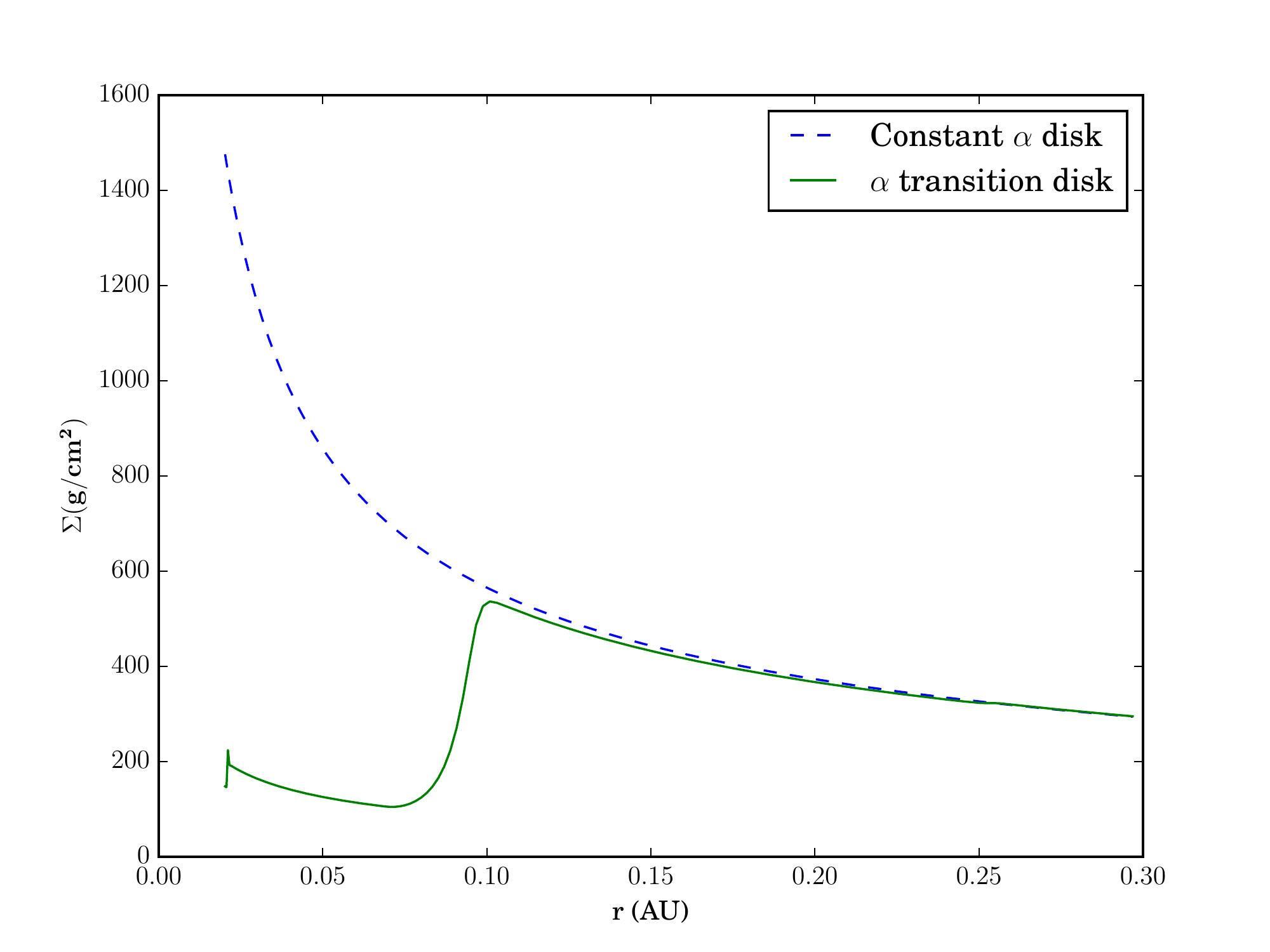} 
\caption{
Mass surface density profile after 1000 orbits of our fiducial
``transition'' disk where $\alpha$ rises from 0.001 at 0.1~AU to 0.01 at
0.07~AU with steady accretion rate of $10^{-9}\:M_\odot\:{\rm
  yr}^{-1}$. The dashed line is the profile of a ``constant $\alpha$'' disk
(i.e., $\alpha=0.001$).}
\label{fig:Sigmaradialprofile}
\centering
\end{figure}

We next set up a disk that has a radial jump in $\alpha$. Moving
inwards, $\alpha$ rises from $\alpha_{\rm DZIB}=0.001$ at 0.1~AU to
$\alpha_{\rm MRI}=0.01$ at 0.07~AU, i.e., over a transition width of
$\Delta r_{\rm DZIB}=0.03$~AU, with the transition described by part
of the function $\alpha_{\rm DZIB} + (1+ {\rm sin} x)(\alpha_{\rm
  MRI}-\alpha_{\rm DZIB})$ from $x=3\pi/2$ (at 0.1~AU) to $x=\pi/2$
(at 0.07~AU). The initial mass surface density profile is also chosen
to give a steady accretion rate across this transition region. The
steady radial profile achieved after 1000 orbits is close to this
initial choice and is shown in Fig.~\ref{fig:Sigmaradialprofile}.

\section{Protoplanet Migration}\label{S:migration}

We study the migration of protoplanets of various fixed masses (0.1,
0.5 and 1.0~$M_G$) that are inserted into the transition zone region
of the disk by measuring the torques exerted on the planet by the
disk. For each planet mass, we run simulations holding the planet at a
constant radius, exploring uniformly from 0.085 to 0.115~AU with a
spacing of 0.001~AU. So for each planet mass there are 31 MR and HR
simulations run, for a total of 186 simulations.  Note we keep the
$\alpha$ viscosity radial profile constant in all the simulations in
this section. We run each simulation for 400 orbits, thus allowing the
gas disk to achieve a quasi equilibrium structure. We then evaluate
the torque on the planet by averaging over the next 100 orbits. Given
the 2D nature of these simulations, we adopt a smoothing length in the
calculation of the planet's gravitational potential of $0.6h$
\citep{2012A&A...541A.123M}.

The torque, $\Gamma$, is scaled as
$\Gamma_0=(M_p/m_*)^2\Sigma_gr^4\Omega_p^2h^{-2}$, where $r$ is the
orbital radius of the planet and $\Sigma_g$ is the average gas mass
surface density at this orbital radius.
The grid size in our standard MR run is more than twice the Hill
radius of a planet with $M_p=0.1M_G$. We therefore carry out HR
simulations with twice greater resolution and examine numerical
convergence.

\begin{figure}
\centering
\plotone{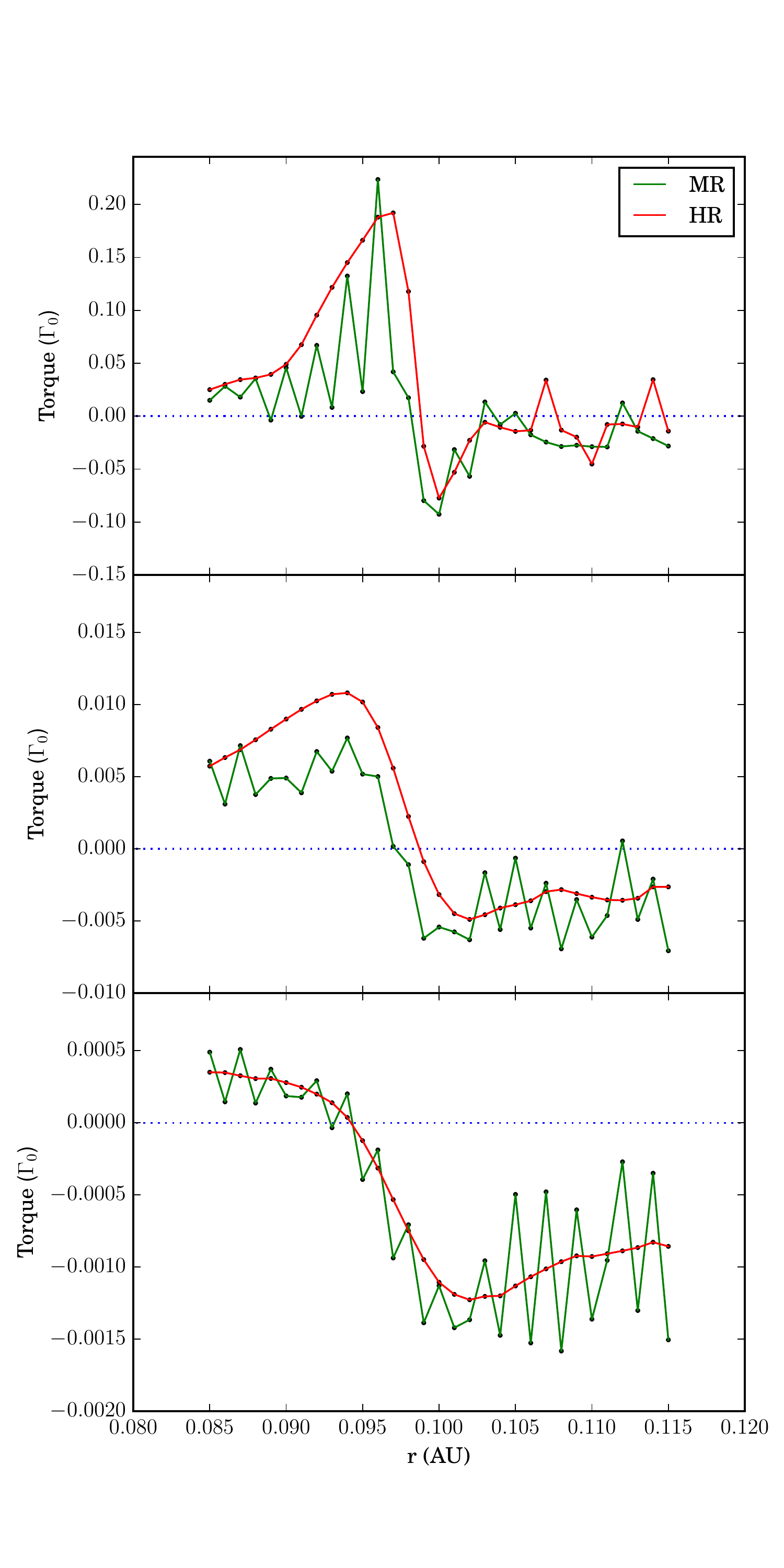}
\caption{
Transition zone migration torque, $\Gamma$, profiles for 0.1~$M_G$
(top), 0.5~$M_G$ (middle) and 1.0~$M_G$ (bottom) planets held at fixed
radius. Medium resolution (MR) and high resolution (HR) runs are
shown, with the torque induced by the disk on the planet being the
average from orbit 400 to 500 after introduction of the planet. A
positive torque leads to outward migration; a negative torque to
inward migration. The HR results show $\Gamma$ decreasing from
positive to negative values in the transition region (where $\alpha$ starts 
decreasing as r increases just interior of 0.1 AU) . 
We expect the planet will migrate inwards from positions where 
the total torque is negative and migrate outwards where the total 
torque is positive. It will stop at zero torque. Thus these torque 
profiles are indicative of there being a stable “planet trap” in this 
location, where $\Gamma=0$.}
\label{fig:planettrap}
\centering
\end{figure} 

In all cases for 0.1~$M_G$, 0.5~$M_G$ and 1.0~$M_G$-mass planets
(which correspond to very different levels of gap opening---see
below), the torque profiles show a common behavior of $\Gamma$
decreasing from positive to negative values in the transition region
(where $\alpha$ starts decreasing as $r$ increases just interior of $0.1$~AU),
indicative of a stable ``planet trap'' in this location, where
$\Gamma=0$ (see Figure~\ref{fig:planettrap}). We note that the
magnitude of $\Gamma$ decreases significantly as $M_p$ increases
from the Type I to Type II regimes.

This confirmation of planet trapping at the DZIB transition zone is a
key requirement of IOPF, since this allows a planet to continue to
grow by pebble accretion from low to relatively high masses. Next we
investigate how pebble accretion may be disrupted by gap opening by
looking at the detailed structure of the gaps.

\section{Gap Opening}\label{S:gapopening}

\subsection{Gap Structure \& Opening Mass for Fixed $\alpha$ Profile}
 
The gap opening process in IOPF is critical to termination of pebble
accretion of the first planet, thus setting its mass. It may also be
important, in combination with dead zone retreat, in setting the
location of the pressure maximum that leads to new pebble ring
formation and then second planet formation.

\begin{figure}
\centering
\includegraphics[width=0.45\textwidth]{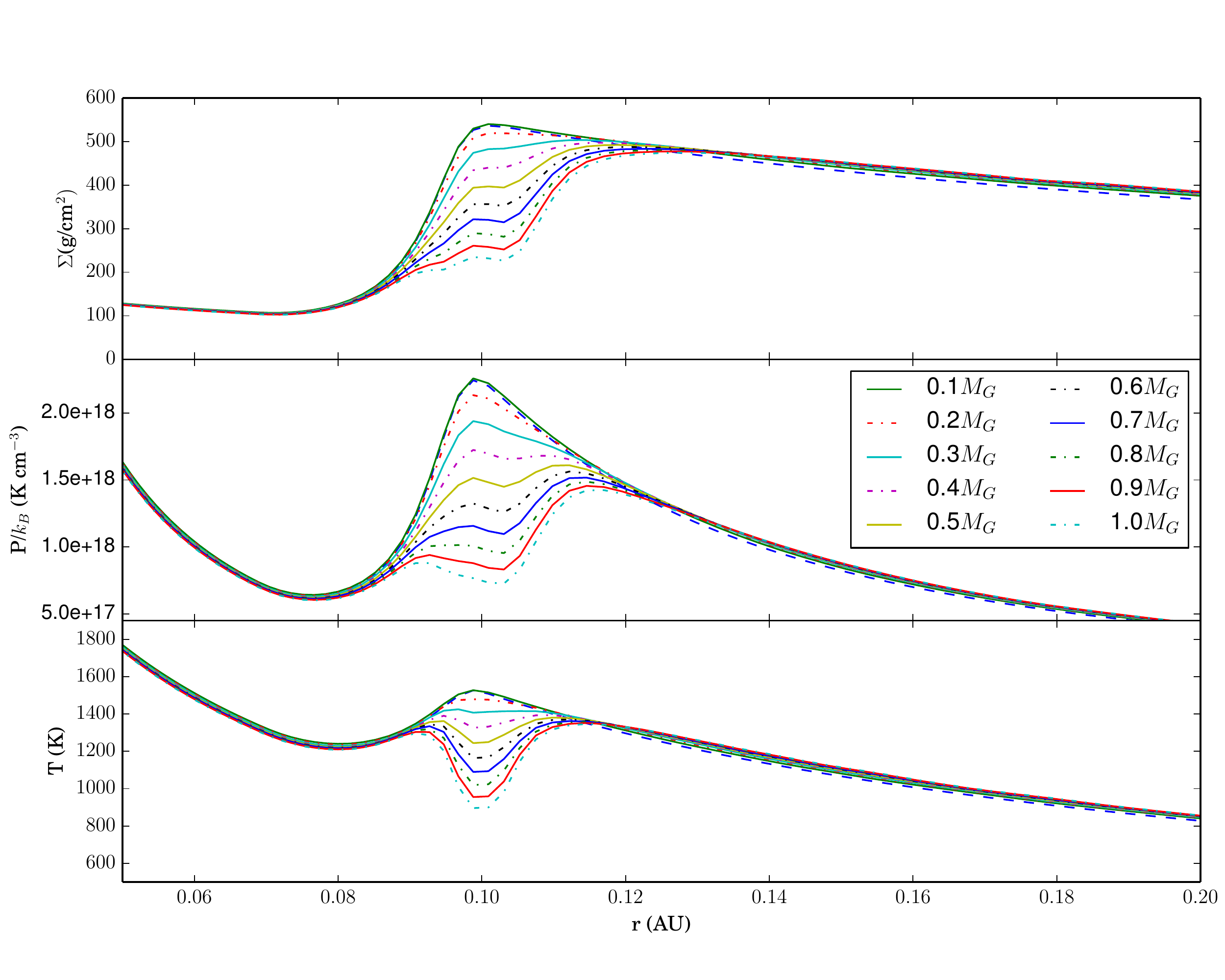}
\caption{
Radial structure of disk mass surface density (top), midplane pressure
(middle) and midplane temperature (bottom) during gap opening for
planets of mass $M_p=0.1\:M_G$ to $1.0\:M_G$ in steps of 0.1~$M_G$
held fixed at the transition zone radius of 0.1~AU. Profiles are shown
for medium resolution runs after 1000 orbits.  Gap opening manifests
itself via a decrease in $\Sigma$, $P$ and $T$ in the vicinity of the
planet and, once $M_p\gtrsim 0.5M_G$, a significant retreat of the
local surface density and pressure maxima outwards by several Hill
radii.}
\label{fig:gap}
\centering
\end{figure}

We now investigate the disk structure that is induced by introducing
planets of various masses at a fixed location of $r=0.1$~AU, i.e., at
the transition zone of the disk with a fixed $\alpha$ profile.
Figure~\ref{fig:gap} shows the radial profiles of mass surface
density, midplane pressure and midplane temperature after 1000 orbits
of evolution from introduction of planets of masses $M_p=0.1\:M_G$ to
$1.0\:M_G$ in steps of 0.1~$M_G$.  We note that the gas profile
settles very quickly, within $\sim 100$ orbits, to a profile close to
that of the final state at 1000 orbits.

As planet mass increases, the radial profiles show a gradual deepening
of the gap in both the mass surface density and pressure profiles. The
temperature also dips due to reduced viscous heating and smaller
optical depths.  For this particular set-up, with a transition zone
width of 0.03 AU, it is only once planet masses are $\gtrsim 0.5 M_G$
 that there is significant displacement of the azimuthally
averaged pressure maximum away from the planet's orbital radius.
Table~\ref{tab:deltar} lists the separation of the pressure maximum
from the planet's orbital radius at 0.1~AU, $\Delta r_{\rm Pmax}$, but
normalized by the planet's Hill radius, $R_H = (M_p/(3m_*))^{1/3}r$.
We define $\phi_{\rm \Delta r,Pmax}\equiv \Delta r_{\rm Pmax}/R_H$. As
$M_p$ increases from 0.4 to 0.5~$M_G$, $\phi_{\rm \Delta r,Pmax}$
grows by a factor of 10 from $\simeq 0.4$ to
$\simeq5$. Figure~\ref{fig:gap} shows this also corresponds to the
mass surface density maximum retreating outwards by a similar amount.

We have compared gap opening results for simulations with $M_p=0.1,
0.5, 1.0 M_G$ at MR and HR resolutions. We find very similar radial
profiles of $\Sigma$ and $P$, with agreement in values at better than
10\% across the width of the gap region.

\begin{figure*}[t]
\centering
\includegraphics[width=1.0\textwidth]{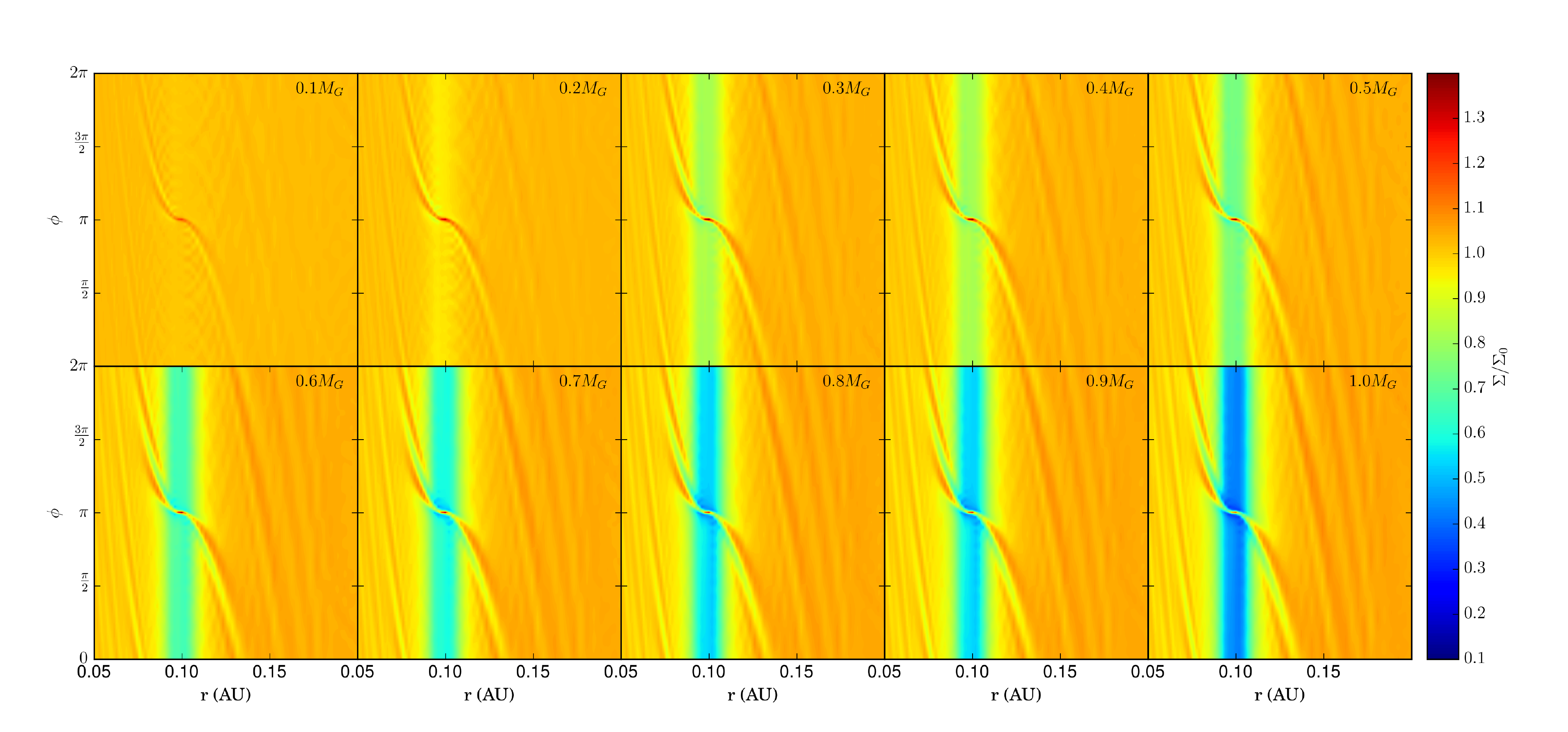}
\caption{
2D $r$-$\phi$ view of the mass surface density perturbation,
$\Sigma/\Sigma_0$, induced by different planet masses from 0.1 to
1.0~$M_G$ (where $\Sigma_0$ is the surface density in the absence of a
planet).
}
\label{fig:2dgap}
\centering
\end{figure*}

Figure~\ref{fig:2dgap} shows the 2D view of the perturbation of the
disk mass surface density (compared to the disk with no planet) in the
$r$-$\phi$ plane. The strengthening spiral arms, along with deepening
gap, are evident.

\begin{table*}[t]
\centering
\caption{Retreat of the local pressure maximum during gap opening}
\begin{tabular}{l|llllllllll}
\toprule
$M_{p}/M_G$ &0.1&0.2&0.3&0.4&0.5&0.6&0.7&0.8&0.9&1.0  \\
\midrule
$\phi_{\rm \Delta r,Pmax}=\Delta r_{\rm Pmax}/R_H$ &0.61&0.48&0.42&0.38&4.75&4.47&5.09&4.86&4.68&5.27 \\
\bottomrule
\end{tabular}
\footnotetext{The separation between new pressure maximum and planet orbital radius, scaled to planet's Hill radius. These results are for disks with fixed $\alpha$ profile with a transition width of $\Delta r_{\rm DZIB}=0.03$~AU. Note, location of pressure maximum is measured at the center of grid cell of maximum pressure. 
}\label{tab:deltar}
\end{table*}

\subsection{DZIB Retreat with an Evolving $\alpha$ Profile}

When a planet opens a gap in the disk leading to decreasing density in
the vicinity of the planet, we expect higher ionization levels given
the $n^2$ dependence of recombination rates, and thus greater
likelihood of activating the MRI, which then results in the retreat of
the dead zone inner boundary. Higher levels of ionization may also
arise due to shock heating in spiral arms that are induced by the
planet. We note that these spiral arms, propagating outwards into the
dead zone, will by themselves also provide a source of enhanced
effective viscosity. The lower densities of the gap region may also
allow for increased penetration of X-rays that raise the ionization
level and activate the MRI. Such effects could cause the viscosity to
make its radial transition further out in the disk, which would lower
the density in the gap region and vicinity even more.

The full treatment of these processes is a very complex problem,
beyond the scope of the present paper. Here we present a simple,
heuristic model for DZIB retreat based on X-ray penetration from the
protostellar corona to the disk midplane. The X-rays are assumed to
propagate from a height of 0.05~AU at $r=0$ (i.e., $11 R_\odot$, which
is expected to be $\sim3$--4 stellar radii for a solar mass pre-main
sequence star). The thickness of the transition zone from the
MRI-active region to dead zone is assumed to be characterized by
$\phi_X\Sigma_X$, i.e., a mass surface density that scales with the
X-ray penetration mass surface density due to absorption, $\Sigma_X$.

Starting from the FARGO simulation with $\phi_G=0.5$ after 300 orbits
(i.e., when a wide, but relatively shallow, gap has just been opened),
the radius of the DZIB, $r_{\rm DZIB}$, is re-evaluated each orbit by
the condition that the mass surface density along the path from the
X-ray corona to the midplane equals $\phi_X\Sigma_X$. For simplicity, we
assume there is negligible material along this path length at
locations interior to the planet, i.e., at $r<0.1$~AU, which we
consider to be a good approximation due to both viscous clearing of
the inner disk and also since the disk's vertical scale height is
small in this region. Thus our path integral starts from $r=0.1$~AU
and extends out to $r_{\rm DZIB}$. Note, that this path starts with a
relatively shallow angle to the midplane of $\simeq24^\circ$, which
then decreases as the disk evolves and $r_{\rm DZIB}$ increases.

In addition, we need to specify the structure of the $\alpha$
transition zone in the X-ray penetration region. We expect it to be
broader than the transition zone set by thermal ionization of alkali
metals. For simplicity, we assume the functional shape of the radial
variation of $\alpha$ is the same as that adopted in
\S\ref{S:numerical}. The outer radius where $\alpha$ starts rising
from the DZIB value of 0.001 is set by the mass surface density along
the pathlength being equal to $\phi_X\Sigma_X$. We note that the width
of this transition region may be somewhat larger than the actual
penetration depth of the X-rays (i.e., $\phi_X>1$), because of, for
example, the outward propagation of turbulence, which has been seen in
the simulations of \citet{2010A&A...515A..70D}.
For the inner radius, $r_i$, where $\alpha$ reaches the full
MRI value of 0.01, we investigate three cases, i.e., inner radii of
0.07~AU (Case A: the same as adopted for our migration and gap opening
studies, above), 0.1~AU (Case B: the location of planet), and 0.11~AU
(Case C: the location of the initial gap outer edge that has been
induced by the planet).

For a simple estimate on the values of $\Sigma_X$ and $\phi_X$ to be
used in this model, we note that X-ray opacity, even at $\sim10$~keV,
may be dominated by heavy elements in the gas, rather than in dust,
especially if dust has mostly coagulated into large grains and pebbles
that have settled to the midplane and been trapped at the DZIB
pressure maximum. For interstellar gas with solar abundances and the
case of heavy elements depletion, \citet{1999ApJ...518..848I} find an
absorption opacity of $\tau_E = 0.056 (\Sigma/1\:{\rm
  g\:cm^{-2}})(E/10\:{\rm keV})^{-2.81}$ (valid for
$E=2$--30~keV). Thus for $\tau_{\rm 10keV} = 1, 10$ requires $\Sigma_X
= 18, 180\:{\rm g\:cm^{-2}}$ and $\tau_{\rm 30keV} = 1, 10$ requires
$\Sigma_X = 39, 390\:{\rm g\:cm^{-2}}$. Thus we choose a
characteristic value of $\Sigma_X=100\:{\rm g\:cm^{-2}}$, equivalent
to absorption optical depths of a few for $\sim 20$~keV X-rays.  We
note, however, that the radiative transfer of these higher energy
X-rays is likely to be controlled by Compton scattering by electrons
present in neutral H, $\rm H_2$ and He, which may allow increased
penetration to the disk midplane at greater radial distances via
scattering from the disk atmosphere. For this reason, along with the
extra thickness of the transition zone that is expected due to
propagation of turbulent wakes from the MRI-active zone (including
vertically from modest scale heights), we choose a fiducial value of
$\phi_X=5$.

From the above discussion, it is clear that more detailed calculations
are needed to better constrain the combined parameter
$\phi_X\Sigma_X$, including ionization equilibrium calculations, e.g.,
similar to those of \citet{1999ApJ...518..848I} and then applying the
resulting resistivities to determine the extent of the MRI-active zone
(Mohanty \& Tan, in prep.). The time variable nature of the X-ray
luminosity may also need to be accounted for.

\begin{figure}
\centering
\includegraphics[width=0.5\textwidth]{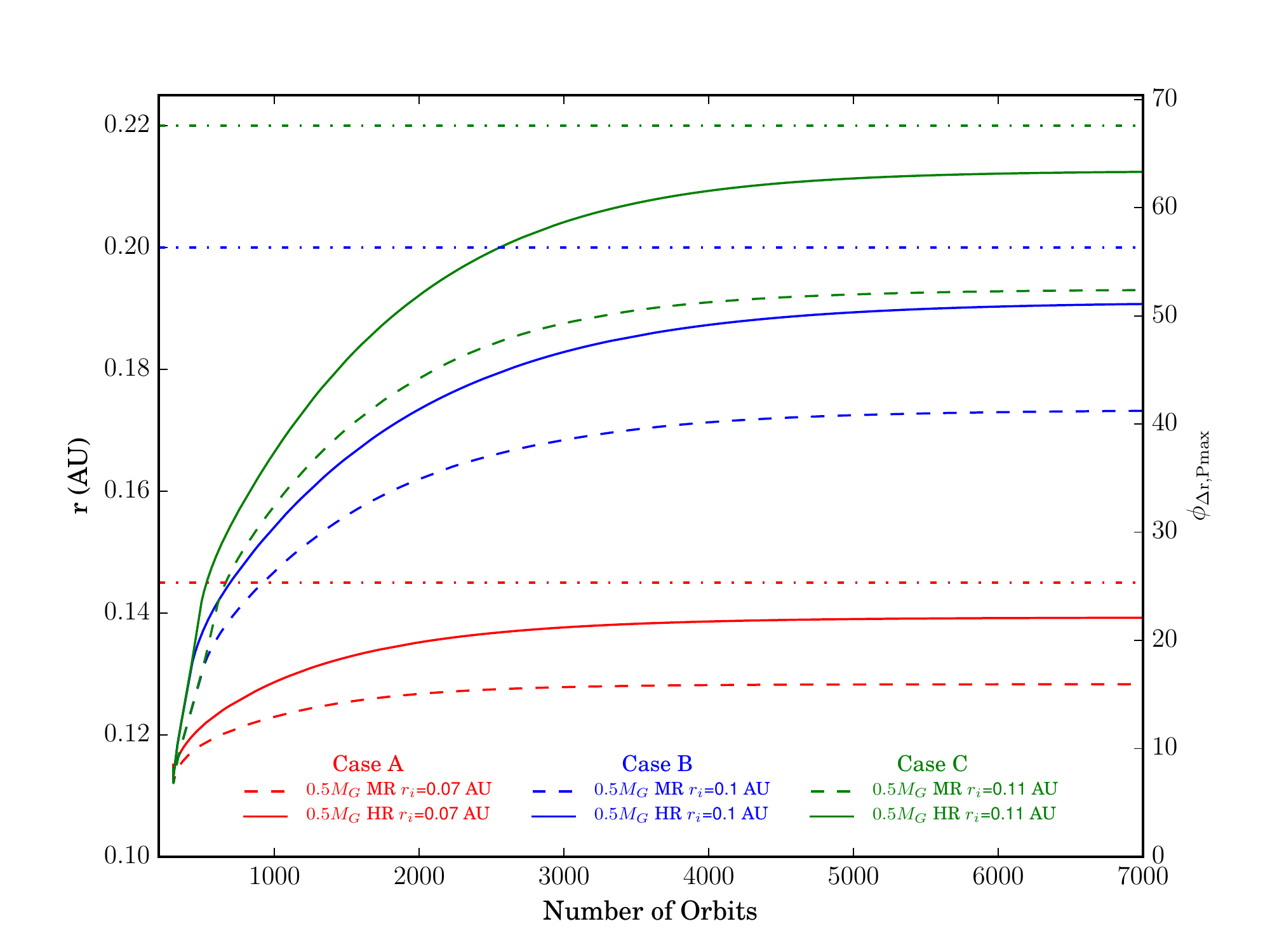}
\caption{
Radius of DZIB (i.e., the outer transition radius of $\alpha$ profile)
as set by X-ray penetration in evolving $\alpha$ disks. Medium
resolution (MR) results are shown with dashed lines; high resolution
(HR) results with solid lines. The three cases of inner transition
radii of 0.07~AU (Case A, red lines), 0.1~AU (Case B, blue lines) and
0.11~AU (Case C, green lines) are shown. In each case, the estimated
converged, final retreat radius is shown with a horizontal dot-dashed
line (see text).}
\label{fig:DZIBretreat}
\centering
\end{figure}

Figure~\ref{fig:DZIBretreat} shows results from these DZIB retreat
models. We find that $r_{\rm DZIB}$ increases gradually over
timescales of several thousand orbits of the planet. For Case A, by
6500 orbits the rate of increase is very slow, i.e., from the 6500th
to 7000th orbit the percentage increase is $<$0.06\% for the MR run
and 0.07\% for the HR run. The equivalent results for Cases B and C
are similar. However, the value of $r_{\rm DZIB}$ after 7000 orbits
does depend somewhat on the resolution of the simulation: for the Case
A MR run $r_{\rm DZIB} = 0.128$~AU; for the HR run $r_{\rm DZIB} =
0.139$~AU, i.e., the amount of retreat ($\Delta r_{\rm Pmax}$) is
about 38.6\% greater in the HR run. We thus also carried out twice
higher, SHR, resolution runs for $\sim$1000 orbits. We find that
$\Delta r_{\rm Pmax}$ is $\lesssim$5\% greater than in the HR
simulation. We use these results to estimate that the final
numerically converged retreat distance is $\sim$5-10\% greater than
the HR result (or about 40\% greater than the MR result), i.e.,
$r_{\rm DZIB} ({\rm Case A}) \simeq 0.145$~AU, i.e., $\Delta r_{\rm
  Pmax} \simeq 0.045$~AU. For this case with
$M_p=0.5M_G=5.59\:M_\oplus$, for which the planet's Hill radius is
$R_H=1.78\times 10^{-3}$~AU then $\phi_{\rm \Delta r,Pmax}\equiv
\Delta r_{\rm Pmax}/R_H = 25.3$, which is a significantly greater
retreat (i.e., $\sim 5\times$ greater) than due to simple gap opening
in the fixed $\alpha$ model.

For Cases B and C the amount of retreat is larger and takes somewhat
longer to stabilize (although the retreat has essentially stopped by
7000 orbits). After 7000 orbits, the HR simulations find a 24.0\% and
20.8\% larger retreat than the MR runs for Cases B and C,
respectively. The SHR results at $\sim 1000$ orbits are 5.6\% and
4.9\% greater, respectively. Thus we estimate final converged values
are about 25\% greater than the MR results, i.e., for Case B DZIB
retreat to 0.20~AU ($\Delta r_{\rm Pmax} \simeq 0.10$~AU; $\phi_{\rm
  \Delta r,Pmax}=56.3$) and for Case C DZIB retreat to 0.22~AU
($\Delta r_{\rm Pmax} \simeq 0.12$~AU; $\phi_{\rm \Delta
  r,Pmax}=67.6$). These results are also summarized in
Table~\ref{tab:variable_deltar}.

As a check on self-consistency, we have also investigated the
potential effect of X-ray penetration inducing an evolving $\alpha$
profile in a disk that does not yet have a planet or gap. For this
model, where the accretion disk extends uninterrupted to very close to
the protostar at $\simeq 0.02$~AU, we set the height of the stellar
X-ray corona to be equal to this inner disk radius, i.e., 0.02~AU. We
also start the path integral for X-ray penetration from this
distance. We find that for the same choices of $\Sigma_X$ and $\phi_X$
adopted above, the X-rays do not penetrate beyond $\sim0.1$~AU and so
the DZIB $\alpha$ profile set by thermal ionization of alkali metals
is not expected to be affected. However, we find that utilizing a
larger X-ray corona height of 0.05~AU does enable X-ray penetration to
beyond 0.1~AU in the no-planet disk. Thus, in the context of this
simple model, DZIB retreat is also influenced by properties of the
scaleheight of X-ray emission from the protostar.

Overall, we therefore see that the amount of DZIB retreat, which in the
IOPF scenario is expected to set the location of second planet
formation, depends somewhat sensitively on the detailed structure of
the $\alpha$ transition zone and the X-ray emission properties of the
protostar. Thus future work to model this zone including the full
physics of activation of the MRI in response to a changing ionization
fraction, e.g., due to X-rays or other ionization sources, is
needed. However, we can still conclude that, for the parameters
explored here, the amount of retreat is significantly greater than the
retreat due to initial gap opening. In \S\ref{S:obs} we compare these
retreat distances to the orbital spacings of the {\it Kepler}-observed
planets.

\begin{table*}[t]
\centering
\caption{DZIB retreat for evolving $\alpha$ models for different inner transition radii, $r_i$}
\begin{tabular}{c|ccc}
\toprule
Models & & $\phi_{\rm \Delta r,Pmax}$  & \\
 & Medium Res.(7000 orbits) & High Res.(7000 orbits) & Estimated Convergence  \\
\midrule
Case A ($0.5M_G$, $r_i$=0.07 {\rm AU}) &15.9 (0.028 AU)&22.1 (0.039 AU)&$\simeq$25.3 (0.045 AU) \\
Case B ($0.5M_G$, $r_i$=0.10 {\rm AU}) &41.2 (0.073 AU)&51.1 (0.091 AU)& $\simeq$56.3 (0.100 AU) \\
Case C ($0.5M_G$, $r_i$=0.11 {\rm AU}) &52.4 (0.093 AU)&63.3 (0.112 AU) & $\simeq$67.6 (0.120 AU)\\
\bottomrule
\end{tabular}
\footnotetext{
$\phi_{\rm \Delta r,Pmax}$ 
measures the separation between the planet (at 0.1 AU) and the outer
transition of $\alpha$ profile (as set by X-ray penetration) in units
of the planet's Hill radius. The outer transition radius is at a
location very similar to that of the local pressure maximum (see
Fig.~\ref{fig:evolalpha}).}
\label{tab:variable_deltar}
\end{table*} 

\begin{figure*}
\centering
\includegraphics[width=1.0\textwidth]{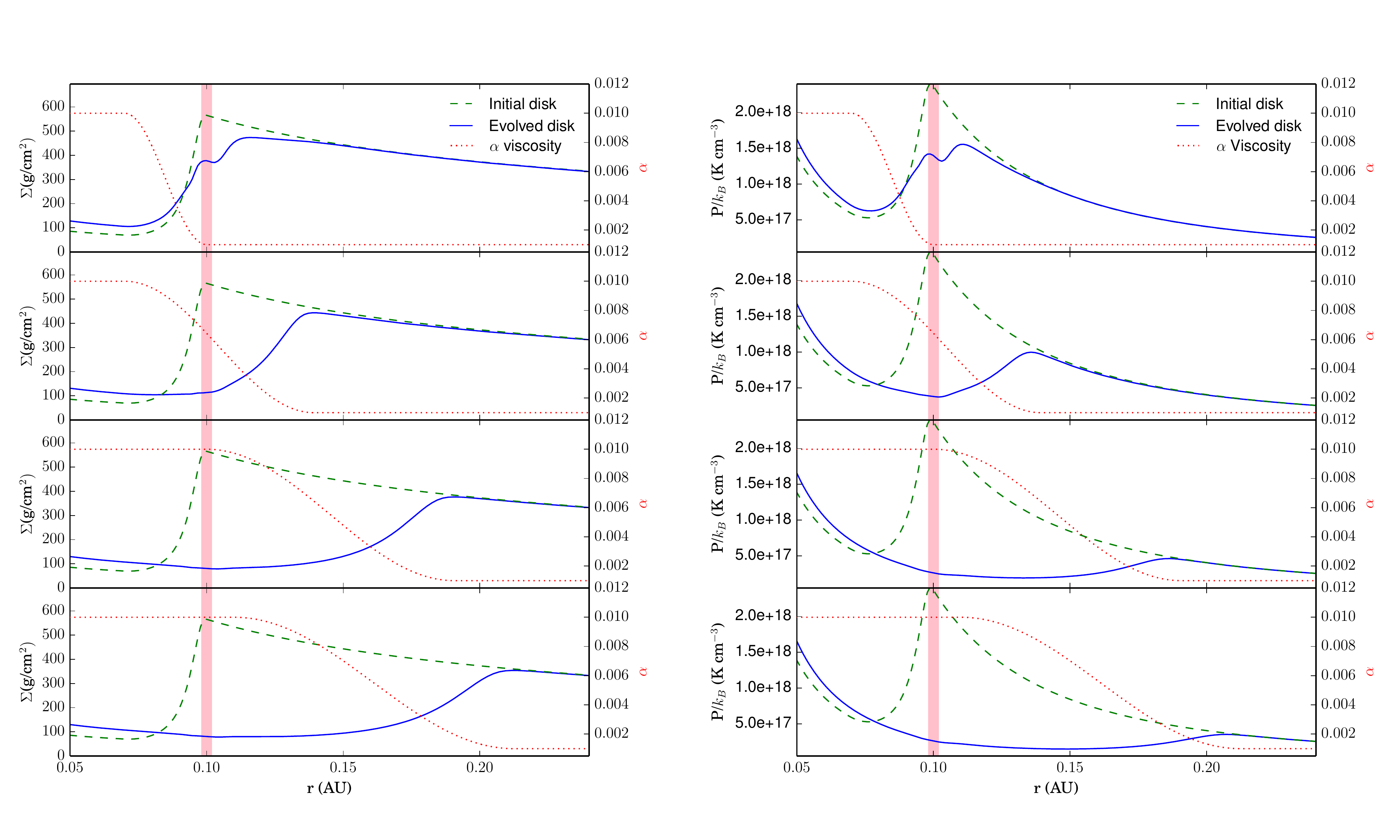}
\caption{
Radial structure ({\it Left panel:} mass surface density; {\it Right
  panel:} midplane pressure) of a disk with a 0.5 $M_G$ planet held
fixed at 0.1~AU. {\it 1st row:} The fixed $\alpha$ profile disk model
($\alpha(r)$ is shown by the dotted red line). Disk structure is shown
after 1000 orbits (solid blue line), compared to initial state (dashed
green line). The extent of the planet's Hill sphere is shown by the
vertical shaded band. {\it 2nd row:} The evolving $\alpha$ disk model
(HR run) at 7000 orbits due to X-ray penetration in which the inner
$\alpha$ transition radius is held fixed at 0.07~AU (Case A). {\it 3rd
  row}: Same as above, but for Case B with inner transition radius at
0.1~AU. {\it Bottom row}: Same as above, but for Case C with inner
transition radius at 0.11~AU.
}
\label{fig:evolalpha}
\centering
\end{figure*}

Figure~\ref{fig:evolalpha} shows the radial structure of the disks for
these evolving $\alpha$ models after 7000 orbits, and compares to
the fixed $\alpha$ profile model, which was already settled after 1000
orbits. This figure shows that the local pressure maximum remains
closely located with the outer $\alpha$ transition radius, i.e., where
it begins to rise from its DZIB value of 0.001.

\begin{figure*}
\centering
\includegraphics[width=1.0\textwidth]{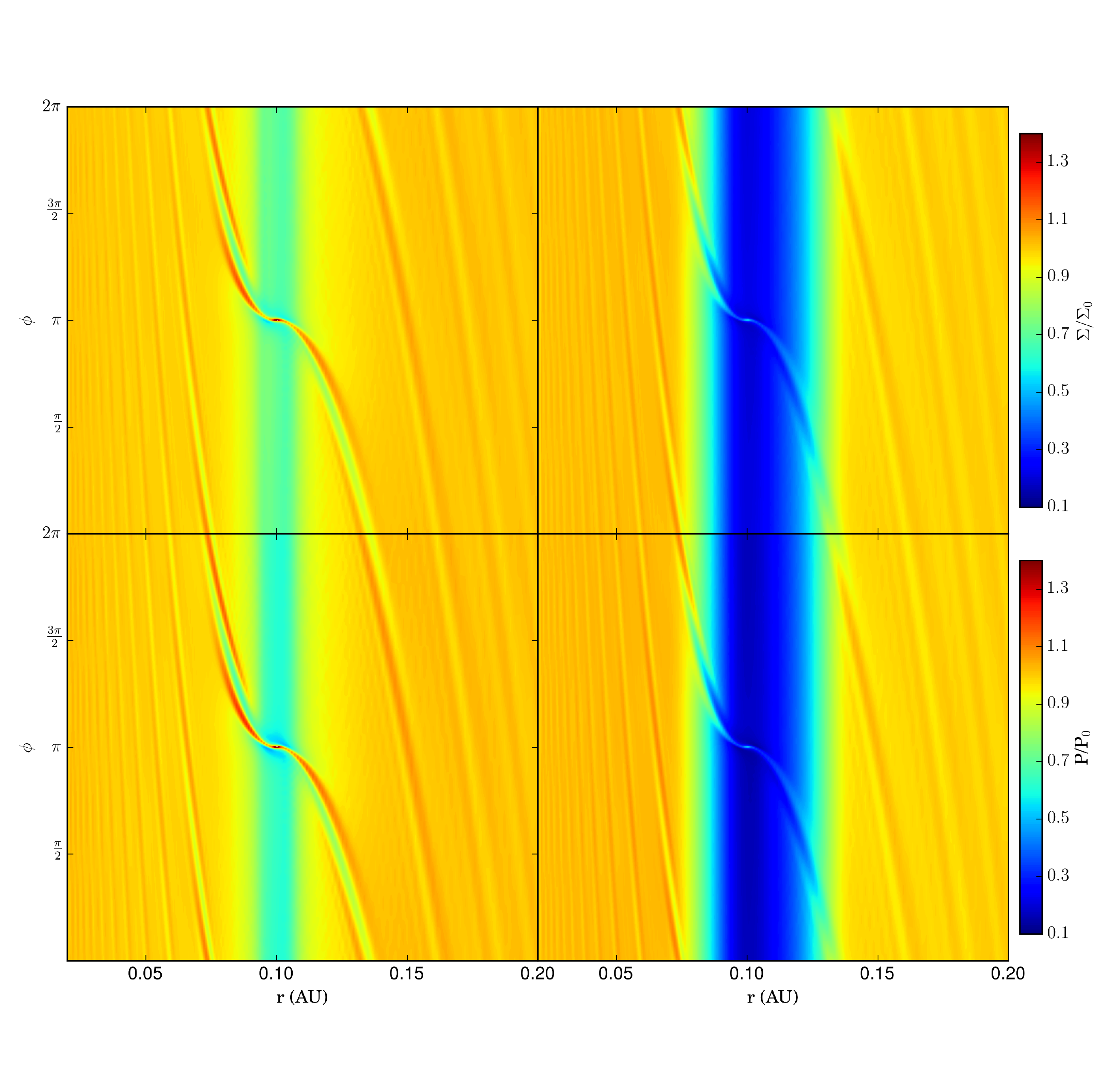}
\caption{
Mass surface density (top row) and midplane pressure (bottom row)
perturbation plots of disks with a $0.5\:M_G$ planet with ({\it Left column}) fixed
$\alpha$ profile (after 1000 orbits) and ({\it Right column}) evolving $\alpha$ profile
(Case A, after 7000 orbits).}
\label{fig:2Devolperturb}
\centering
\end{figure*}

\begin{figure*}
\centering
\includegraphics[width=1.0\textwidth]{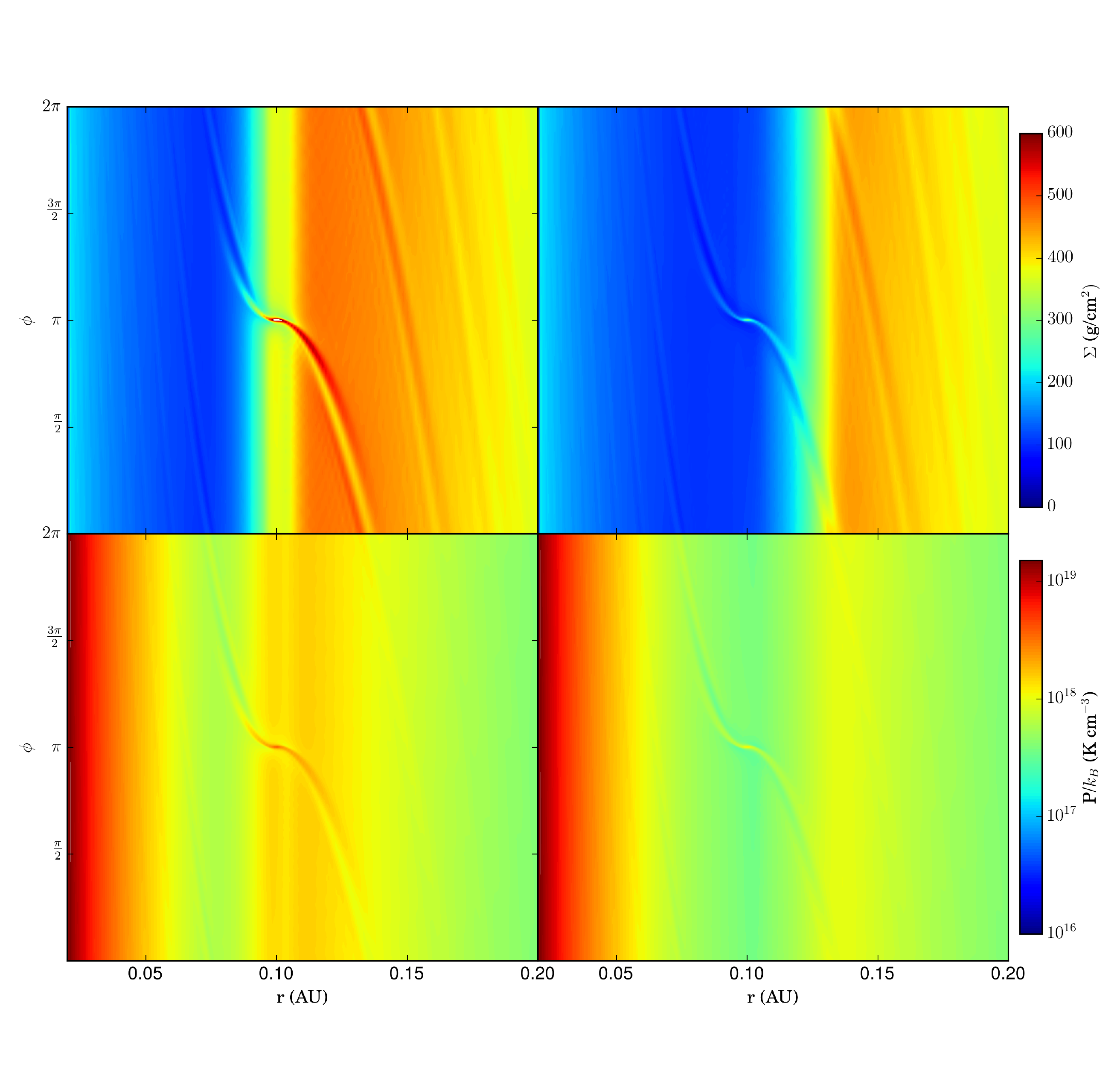}
\caption{
Same as Figure~\ref{fig:2Devolperturb}, but now showing absolute
values of mass surface density (top row) and midplane pressure (bottom
row) of disks with a $0.5\:M_G$ planet with ({\it Left column}) fixed
$\alpha$ profile (after 1000 orbits) and ({\it Right column}) evolving
$\alpha$ profile (Case A, after 7000 orbits).}
\label{fig:2Devol}
\centering
\end{figure*}

\begin{figure*}
\centering
\includegraphics[width=1.0\textwidth]{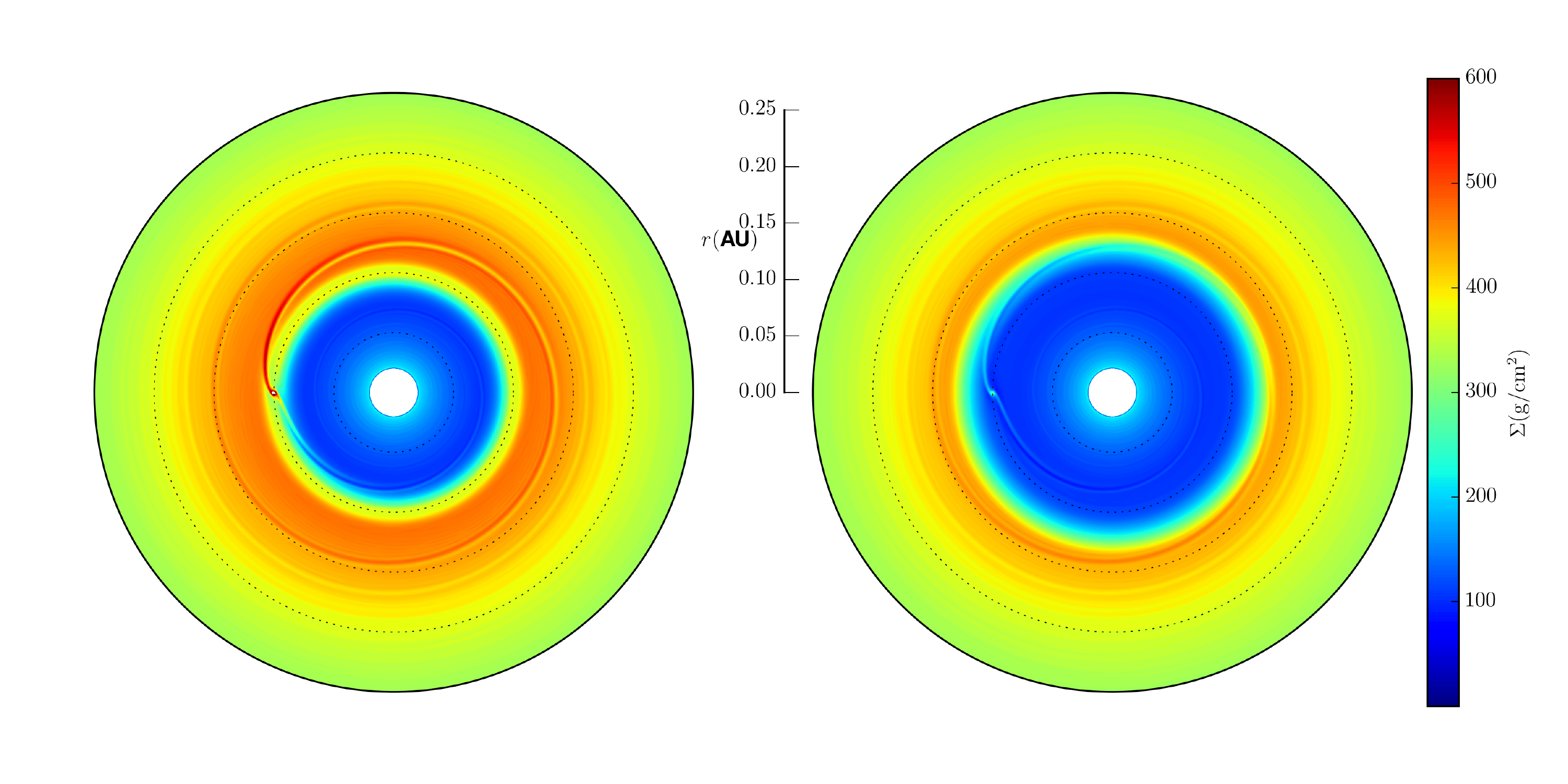}
\caption{
Same as Figure~\ref{fig:2Devol} top row, but now showing a spatial map
of the disk structure induced by a $0.5\:M_G$ planet with ({\it Left}) fixed
$\alpha$ profile (after 1000 orbits) and ({\it Right}) evolving
$\alpha$ profile (Case A, after 7000 orbits).}
\label{fig:2Devolpolar}
\centering
\end{figure*}

Figure \ref{fig:2Devolperturb} shows the 2D $r$-$\phi$ plots of
perturbations of mass surface density and pressure for the Case A
evolving $\alpha$ model and compares to the fixed $\alpha$ model, both
of which have $M_p=0.5M_G$. Figure~\ref{fig:2Devol} shows the absolute
values of these quantities. For completeness, to better illustrate
physical appearance of disk structures, we also show true spatial maps
of disk mass surface density structure of Case A in
Figure~\ref{fig:2Devolpolar}. These figures reveal the decreasing
absolute mass surface density of the induced spiral arm in the case
where DZIB retreat has proceeded further.

We emphasize that we have presented a very simple, heuristic model for
a disk in which the viscosity profile responds actively to X-ray
penetration (which itself is sensitive to the disk's structure). More
accurate modeling would involve self-consistent calculation of the
viscosity caused by the MRI in the transition zone
\citep[e.g.,][]{2011ApJ...736..144B,2013ApJ.764.65}, including
calculation of the ionization structure of the disk as it is affected
by gap opening. The results that are needed from such modeling are the shape
and absolute values of the disk viscosity in the MRI transition zone,
which could then be implemented parametrically in, e.g., FARGO
hydrodynamic simulations of the disk. However, potential instability
of the dead zone inner edge, which may lead it to vary its radial
location \citep{2012MNRAS.424.1977L}, or vortex formation at the inner
edge \citep{2015A&A...573A.132F}, should also be assessed. Although
the model presented so far should be regarded as illustrative, as we
discuss in \S\ref{S:obs}, the scale of pressure maximum retreat that
it exhibits is comparable to the orbital separations of the innermost
planets in STIPs.

\section{Implications for Observed Planets}\label{S:obs}

CT14 examined the separation of planet orbits in STIPs (normalized by
Hill radius of inner planet in the pair), finding a broad range of
separations from innermost, Vulcan, planet to next planet of
$\phi_{\Delta r,1}\sim$few--100, with a broad peak from $\sim
20-60$. Note, however, that this analysis was based on the simple,
single power law planetary mass-size relation of
\citet{2011ApJS..197....8L}. 

\begin{figure*}[t]
\centering
\plotone{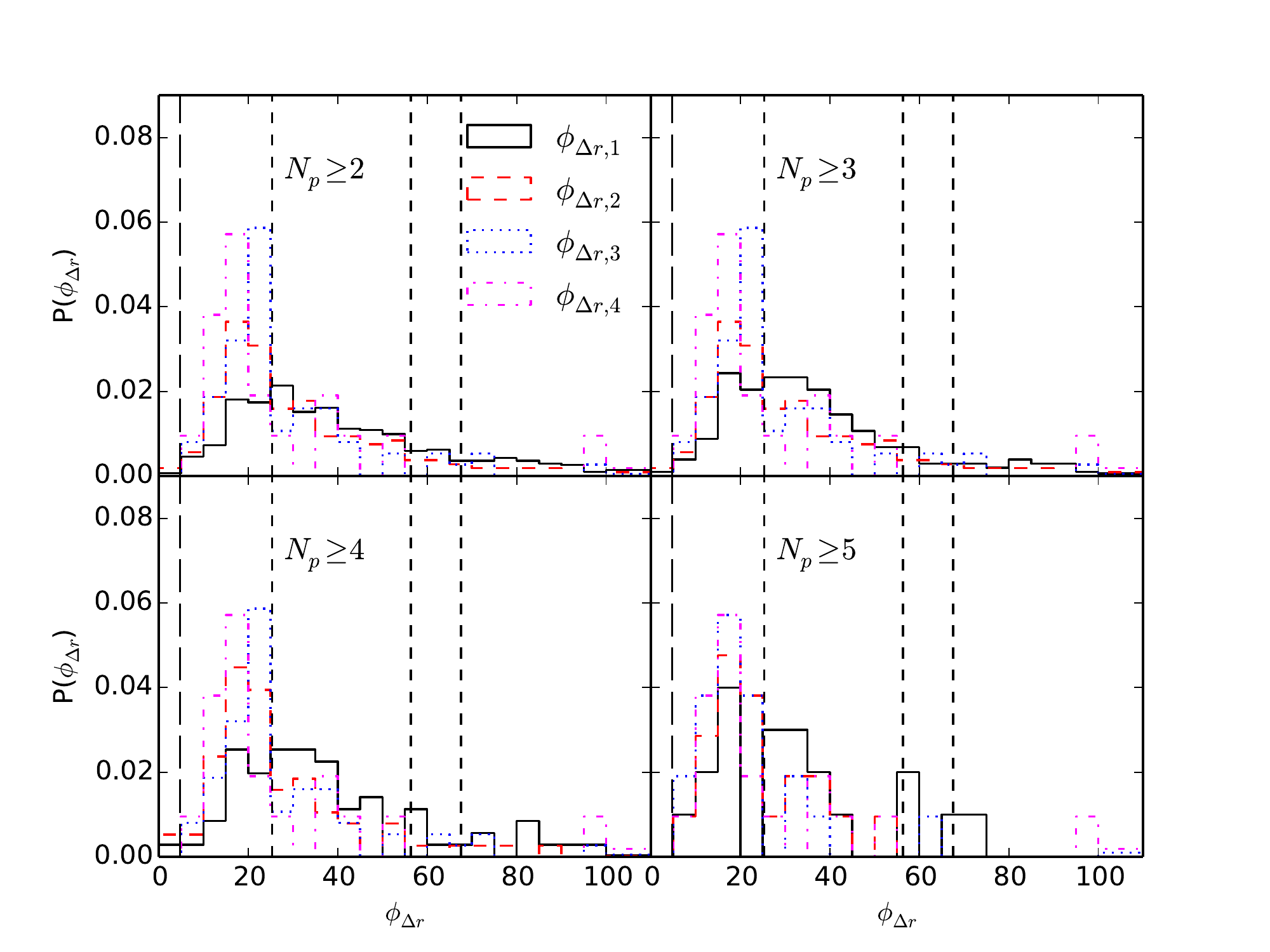}
\caption{
A comparison of orbital separations of the {\it Kepler}-detected
planets (see legend) and our simulation results for DZIB pressure
maximum retreat: vertical long-dashed line shows $\phi_{\Delta r, \rm
  Pmax} = 4.75$ for the initial gap-openning in the fixed $\alpha$
profile disk. The three short dashed lines show the estimated values
of $\phi_{\Delta r, \rm Pmax} = 25.3, 56.3, 67.6$ for Cases A, B, C,
respectively, of the evolving $\alpha$ profile disk.}
\label{fig:obs}
\centering
\end{figure*}

CT15 analyzed the available mass-size data for STIPs planets and
derived an improved, piece-wise power law mass-size relation. Here, we
use this relation (PL3 in CT15) to re-evaluate the distributions of
$\phi_{\Delta r,1}$ through $\phi_{\Delta r,4}$, where the numerical
value of the subscript refers to the number of the planet that is the
innermost in the pair being considered, counting out from the
star. These results are presented in Figure~\ref{fig:obs}.

Note some of the dispersion in separations is likely to be induced by
an intrinsic dispersion in the densities of the planets, leading to
inaccurate estimation of masses. For example, the observed dispersion
in density of a factor of about 5 (e.g., CT15) leads to a dispersion
in inferred mass of the same factor and a dispersion in $R_H$ and
$\phi_{\Delta r}$ of a factor of 1.7. Also, the observed values of
$\phi_{\Delta r,1}$ may be overestimated if some fraction of second
planets are non-transiting and thus not detected due to slight orbital
misalignments.

In Figure~\ref{fig:obs} we also show the locations of $\phi_{\rm
  \Delta r,Pmax}$ for the fixed and evolving $\alpha$ disk models we
considered. We note that the evolving $\alpha$ disk models can achieve
separations of $\sim 20$ to 70 $R_H$, which overlap with the
observed distribution of $\phi_{\Delta r,1}$.

As discussed by CT14, the distributions of $\phi_{\Delta r,1}$ are
significantly different from those of the outer separations. This
conclusion remains unchanged by our use of the improved piecewise
power-law CT15 mass-size relation. For KPC systems with $N_p \geq 3$,
the probability that $\phi_{\Delta r,1}$ and $\phi_{\Delta r,2}$ are
drawn from the same distribution is $9\times 10^{-4}$.  Similarly, for
systems with $N_p\geq4$, the respective probabilities that two
$\phi_{\Delta r}$ distributions are drawn from the same underlying
distribution are $7\times 10^{-5}$, $2\times 10^{-6}$, and $0.36$ for
$\phi_{\Delta r,1}$ and $\phi_{\Delta r,2}$, $\phi_{\Delta r,1}$ and
$\phi_{\Delta r,3}$, and $\phi_{\Delta r,2}$ and $\phi_{\Delta
  r,3}$. This indicates that the distribution of separations between
the first two planets is different than between the 2nd and
3rd and other outer pairs of planets in these systems.

As also discussed by CT14, the generally larger values of $\phi_{\Delta
  r,1}$ may be a signature of the greater relative effect of inner
disk clearing after first, Vulcan planet formation on the location of
the second planet, compared to later, more incremental dead zone retreats.

\section{Summary and Discussion}\label{S:conclusions}

We have explored three main aspects of the Inside-Out Planet Formation
(IOPF) scenario with numerical simulations of planet-disk
interactions. Our main findings are the following:

(1) Planets of all relevant masses are trapped at the dead zone inner
boundary transition region, so planetary migration (of Type I or Type
II) does not appear to be a problem for IOPF. Especially during the
early stages of potential Type I migration, the fact that the
protoplanet is trapped at its formation location should allow it to
continue to accrete, especially by pebble accretion. 

(2) The mass scale of significant gap opening that can affect pebble
accretion by displacing the local pressure maximum away from the
planet is $M_p\simeq 0.5 M_G$, i.e., $\phi_G\simeq 0.5$, which is
about $6\:M_\oplus$ in the fiducial case at $r=0.1~AU$. This is
similar to the fiducial value adopted by CT14, especially since given
the renormalization of $M_G$ by a factor of 0.745 in this paper (\S2),
this means that the overall change in planet mass is only a factor of
1.24. However, these results are derived in the context of pure
hydrodynamic simulations and may be sensitive to including extra
physics of MRI activation in the gap region.

(3) A simple model for MRI activation in the gap region, implemented
by an evolving $\alpha$ viscosity profile set by X-ray penetration,
leads to greater retreat of the pressure maximum out to separations of
$\sim 20$ to 70 Hill radii of the planet, but sensitive to model
assumptions about the shape of the viscosity profile in the MRI-active
region to dead zone transition region. Such separations overlap with
the observed orbital separations of innermost planets in STIPs. Future
work is needed to better model viscosity profiles in such transition
regions.

While the above results are supportive of the IOPF model for being
relevant for formation of STIPs, a number of open questions
remain. For example, even though planets with masses $\gtrsim 0.1 M_G$
appear likely to be trapped at a location where they can continue to
grow by pebble accretion, it is possible that this is not the case at
earlier stages for lower mass protoplanets. \citet{2015A&A...573A.132F} have
pointed out that inwardly migrating vortices can form at the dead zone
inner edge and they can potentially interact with the planet, possibly
causing it to also migrate inwards. However, their 3-D simulation
results are for an unstratified disk and the mass accretion rate is
not a constant across the dead zone inner edge, leading to mass pile
up and potentially influencing vortex formation. The potential effects
of vortex interactions still needs to be confirmed in 3-D stratified
simulations where the presence of the surface active layer may
transport the disk mass smoothly across the dead zone inner edge. In
our 2-D simulations, we used a small viscosity in the dead zone region
to represent the effects of an active layer and/or turbulent wakes
spreading from the inner MRI-active region, which may control the
accretion rate in this radial region of the disk. With this set-up we
do not observe the sharp density peaks at the dead zone inner edge
that were present in the \citet{2015A&A...573A.132F} simulations.

Further work is needed on the IOPF model to study the transition of
the pebble ring to a single dominant protoplanet, potentially
involving streaming and gravitational instabilities of the pebble
population and/or Rossby wave instabilities in the gas, followed by
oligarchic growth of planetesimals combined with continued pebble
accretion. Improved study of pebble supply truncation during the
process of gap opening, including utilizing 3-D simulations
\citep[see][for example calculations focussed on the outer
  disk]{2014A&A...572A..35L}, are also needed.

Potential migration of planets after their main accretion phase and
retreat of their natal DZIB, including due to interactions between
neighboring planets, is another topic for future study. However, we
note that for the fiducial IOPF model, the gas mass that is present in
the vicinity of first planet to form is relatively small compared to
the planet's mass, thus limiting the scope of its migration.

\acknowledgements We thank F. Masset and S. Mohanty for helpful
discussions, as well as the comments of an anonymous referee. We
acknowledge support from NASA ATP grant NNX15AK20G (PI: JCT). The
authors acknowledge University of Florida Research Computing for
providing computational resources and support that have contributed to
the research results reported in this publication.

\appendix

\section{Derivation of disk properties}\label{app:equations}

Here we update the analysis presented in CT14, which involves modest
changes in disk properties, as discussed in \S\ref{S:analytic}.
Conservation of angular momentum in a viscous disk implies:
\begin{equation}
r {\frac{\partial }{\partial t} }\left(\Sigma_g r^2 \Omega \right) +\frac{\partial}{\partial r}(r\Sigma_g r^2\Omega v_r)=\frac{1}{2\pi}\frac{\partial \Gamma}{\partial r},
\label{angular momentum}
\end{equation}
where $\Gamma(r,t)$ is the shear torque from viscosity:
\begin{equation}
\Gamma(r,t)=2\pi r \nu \Sigma_g r^2 \frac{d \Omega}{dr}.
\end{equation}
In a steady disk, we have ${\frac{\partial }{\partial t}
}\left(\Sigma_g r^2 \Omega \right)=0$, so integrating \ref{angular
  momentum} we obtain:
\begin{equation}
r\Sigma_g r^2 \Omega v_r = \frac{\Gamma}{2\pi}+C.
\end{equation}
Substituting $\Gamma(r,t)$, we obtain:
\begin{equation}
\nu\Sigma_g\frac{d\Omega}{dr}+\frac{C}{r^3}=\Sigma_g\Omega v_r.
\label{sigma_omega}
\end{equation}
The boundary condition at $r=r_*+\delta r$, $\frac{\partial
  \Omega}{\partial r}=0$, so constant $C$ is:
\begin{equation}
C=r^3\Sigma_g v_r \Omega(r_*+\delta r)=-\dot{m}(Gm_*r_*)^{1/2}/(2\pi),
\end{equation}
where
\begin{equation}
\dot{m}=2\pi r\Sigma_g(-v_r).
\end{equation}

For $\delta r \ll r_*$, we have $\Omega \left(r_*+\delta r
\right)\approx \Omega_K(r_*)=\sqrt{{Gm_*}/{r_*^3}}$, and substituting
into \ref{sigma_omega}, we have:
\begin{equation}
\nu\Sigma_g=\frac{\dot{m}}{3\pi}\left[1-\left(\frac{r_*}{r}\right)^{1/2}\right].
\label{nu_sigma}
\end{equation}
So the energy dissipation rate per unit area (only via one face of the
disk) at radius r is:
\begin{equation}
D(r)=\frac{\Gamma (d\Omega/dr)}{4\pi r}=\frac{\nu\Sigma_g r^2}{2}\left(\frac{d\Omega}{dr}\right)^2=\frac{3Gm_*\dot{m}}{8\pi r^3}\bigg[1-\left(\frac{r_*}{r}\right)^{1/2}\bigg].
\end{equation}

From the mid-plane to disk surface, the optical depth is:
\begin{equation}
\tau_{\rm tot}=0.5\Sigma_g \kappa
\end{equation}

From Eq.~3.11 in \citet{1990ApJ...351..632H}, we know the relation between surface
temperature and actual temperature where the optical depth is $\tau$
to the surface of the disk:

\begin{equation}
T^4 (\tau) =\frac{3}{4}T^4_{\rm eff}\tau\left(1-\frac{\tau}{2\tau_{\rm tot}}\right)=\frac{3}{8}T^4_{\rm eff}\tau_{\rm tot}.
\end{equation}
Here $\tau_{\rm tot}=0.5\Sigma\kappa$, and in our situation, $D(r)=\sigma T^4_{\rm eff}$, $T_c=T\left(\tau_{\rm tot}\right)$, so
\begin{equation}
\frac{16\sigma}{3\Sigma\kappa}T^4_c=D(r).
\end{equation}
We have an expression for mid-plane temperature:
\begin{equation}
T^4_c=\frac{9Gm_*\dot{m}\Sigma_g\kappa}{128\pi\sigma r^3}\left[1-\left(\frac{r_*}{r}\right)^{1/2}\right],
\label{T_c4}
\end{equation}
but we still have an unknown $\Sigma$ profile, so now we try to deduce it. In an $\alpha$ disk, we have:
\begin{equation}
\nu=\alpha \frac{c_{s}^2}{\omega_K}.
\end{equation}
Substituting in \ref{nu_sigma}, we obtain:
\begin{equation}
\Sigma_g =\frac{\dot{m}\sqrt{Gm_*}}{3\pi\alpha c_{s}^2 }\left[1-\left(\frac{r_*}{r}\right)^{1/2}\right]r^{-3/2}.
\end{equation}
For an adiabatic sound speed $c_{s}$, we have:
\begin{equation}
c_{s}^2=\frac{dP}{d\rho}=\gamma\frac{P}{\rho}.
\end{equation}
Then the mid-plane pressure satisfies:
\begin{equation}
P=\frac{\rho k T_c}{\mu}+\frac{4\sigma}{3c} T_c^4\approx \frac{\rho k T_c}{\mu}.
\end{equation}
Now we have the following relation between $\Sigma_g$, $r$ and $T_c$:
\begin{equation}
\Sigma_g=\frac{\dot{m}\sqrt{Gm_*}}{3\pi \alpha}\frac{\mu}{\gamma k T_c}\left[1-\left(\frac{r_*}{r}\right)^{1/2}\right]
r^{-3/2},
\end{equation}
so, substituting for $T_c(r)$ from equation \ref{T_c4}, the surface density profile is:
\begin{equation}
\Sigma_g=\frac{2^{7/5}}{3^{6/5}\pi^{3/5}}\left(\frac{\mu}{\gamma k_B}\right)^{4/5}\left(\frac{\kappa}{\sigma_{\rm SB}}\right)^{-{1}/{5}} \alpha^{-4/5}\left(Gm_*\right)^{1/5}\left(f_r\dot{m}\right)^{3/5}r^{-{3}/{5}},
\end{equation}
and mid-plane temperature (here we use $T$ instead of $T_c$) is:
\begin{equation}
T=\frac{3^{1/5}}{2^{7/5}\pi^{2/5}}\left(\frac{\mu}{\gamma k_B}\right)^{1/5}\left(\frac{\kappa}{\sigma_{\rm SB}}\right)^{{1}/{5}} \alpha^{-1/5}\left(Gm_*\right)^{3/10}\left(f_r\dot{m}\right)^{2/5}r^{-{9}/{10}}.
\end{equation}
For scale height $h$, usually it is assumed that the disk is
vertically isothermal in a passive disk, here in an active disk, we
use an adiabatic vertical structure:
\begin{equation}
\frac{h}{r}=\frac{c_s}{v_\phi}=\frac{c_s}{v_K}=\frac{c_s}{\sqrt{Gm_*/r}},
\end{equation}
so the aspect ratio is:
\begin{equation}
\frac{h}{r}=\frac{3^{1/10}}{2^{7/10}\pi^{1/5}}\left(\frac{\mu}{\gamma k_B}\right)^{-2/5}\left(\frac{\kappa}{\sigma_{\rm SB}}\right)^{{1}/{10}} \alpha^{-{1}/{10}}\left(Gm_*\right)^{-{7}/{20}}\left(f_r \dot{m}\right)^{1/5}r^{{1}/{20}}.
\end{equation}
Since the disk vertical structure satisfies $\rho(z)=\rho(z=0)
e^{-z^2/2h^2}$, mid-plane density $\rho=\Sigma_g/\sqrt{2\pi}h$ can be expressed as:
\begin{equation}
\rho=\frac{2^{8/5}}{3^{13/10}\pi^{9/10}}\left(\frac{\mu}{\gamma k_B}\right)^{6/5}\left(\frac{\kappa}{\sigma_{\rm SB}}\right)^{-{3}/{10}} \alpha^{-7/10}\left(Gm_*\right)^{11/20}\left(f_r\dot{m}\right)^{2/5}r^{-{33}/{20}}.
\end{equation}
\clearpage
\end{CJK*}
\end{document}